\documentclass[aps,prl,twocolumn,showpacs,english]{revtex4}
\usepackage[dvips]{graphicx}
\usepackage{babel} 
\usepackage{color}
\usepackage{ifthen}  
\usepackage{longtable}
\usepackage{natbib}
\usepackage{dcolumn}
\usepackage{amssymb,amsmath}
\usepackage{axodraw4j}
\usepackage{pstricks}
\def\Tr{\text{Tr}}
\def\la{\langle}
\def\ra{\rangle}
\def\be{\begin{equation}}
\def\ee{\end{equation}}
\def\bea{\begin{eqnarray}}
\def\eea{\end{eqnarray}}
\def\bean{\begin{eqnarray*}}
\def\eean{\end{eqnarray*}}
\def\hpsid{{\hat \psi} ^{\dag}}
\def\hpsi{\hat \psi}
\def\psid{\psi^{\dag}}
\def\x{{\bf x}}

\def\y{{\bf y}}

\def\ve{\varepsilon}
\def\3di{{\cdot}}
\def\4di{{\scriptstyle \circ}}

\def\PH{{\mathcal G}}

\def\lett{Letter} 
\begin{document}
\addtolength\topmargin{0.08in}
\renewcommand{\baselinestretch}{1.0}

\draft
\title{The Density Functional via Effective Action}

\date{August 9th, 2009 }

\author{Yi-Kuo Yu}
\affiliation{National Center for Biotechnology Information, National Library of Medicine\\
National Institutes of Health, Bethesda, MD 20894, USA}
\begin{abstract}
A rigorous derivation of the density functional via the effective action in the Hohenberg-Kohn
theory is outlined.  Using the auxiliary field method, in which the electric coupling constant
$e^2$ need not be small, we show that the loop expansion of the exchange-correlation functional 
\mbox{can be reorganized} so as to be expressed entirely in terms of the Kohn-Sham single-particle orbitals and energies. 
\end{abstract}

\maketitle

\renewcommand{\baselinestretch}{0.98}
Interactions among electrons largely determine the structure, 
 phases, and stability of matter. \mbox{Pragmatic} advances in this subject, however, are
 nontrivial. When the number of electrons involved becomes large, calculations 
 based on constructing many-electron wave functions soon lose accuracy and
  will be stopped by an ``exponential wall"\cite{Kohn_99}.   
Density functional theory (DFT), using the three-dimensional electronic
 density as the basic variable, is free from this wall. 
 DFT originated from the theorem of
Hohenberg and Kohn (HK)\cite{HK_64}, which states that there exists a {\it unique}
description of a many-body system in its ground state 
 in terms of the expectation value of the particle-density operator. 
 The HK theorem assures that the ground state energy $E_g$ is obtained 
  by minimizing the energy functional $E_\upsilon$ with respect to the electronic density $n$:
 \vspace*{-4pt}
\be
E_g=\min\limits_n E_\upsilon \left[ n\right] .  \label{enfunc}
 \vspace*{-4pt}\phantom{12}
\ee
 Mermin~\cite{Mermin_65} extended this theorem to  
 finite-temperature.

To make practical use of the HK theorem, a suitable computational scheme
 is necessary. Kohn and Sham~\cite{KS_65} 
proposed a decomposition scheme, aiming to 
 express $E_\upsilon[n]$ via an auxiliary, {\it noninteracting} 
 system that yields a particle density identical to that of the physical ground state. For a 
  nonrelativistic  fermion system described by  
\bea
\hat H &=&\int d{\bf x}\hpsid ({\x})\left( -\frac 1{2m}%
\nabla ^2+\upsilon ({\x}) - \mu \right) \hpsi ({\x})  \nonumber \\
&&\ +\frac{e^2}2\int \int \frac{\hpsid({\x})\hpsid({\y})\hpsi ({\y})\hpsi ({\x})}{|{\x}-{\y}|%
}d{\x}d{\y},  \label{Ho} 
\eea
the energy functional, with 
$e^2$ representing the electric coupling constant and
 $T_0[n]$ being the kinetic energy of the auxiliary system, takes the form 
\bea
E_\upsilon \left[ n\right] &=&\int \upsilon (\x )\, 
n\left( \x\right) d{\x} - \mu N_e  + T_{0}\left[ n\right] \nonumber \\
&& +\frac{e^{2}}{2}\int \int \frac{n({\x})n({\y})}{|{\x}-{\y}|}d\x d\y 
+ E_{xc}\left[ n\right] ,   \label{KSfunctional}
\eea
where $\mu =$ chemical potential, $N_e=$ number of electrons, $\upsilon(\x)=$ external potential, 
and $E_{xc}\left[ n\right] $ is the so-called exchange-correlation energy functional.
 This {\it exact} decomposition cannot exist without the quantity $\frac{\delta E_{xc}[n]}{\delta n}$ being well defined.  
 Being independent of  $\upsilon (\x)$, the sum of the last three terms in (\ref{KSfunctional}) 
  is universal.  All of the many-particle complexity 
  is now completely hidden in $E_{xc}\left[ n\right] $.  

Although $T_0[n]+ E_{xc}[n]$ admits no free parameter and is universal~\cite{HK_64}, its
 explicit construction remains elusive, and parameter-containing {\it empirical} functionals
 are therefore introduced.  
  Cases of failure and limitations of these empirical functionals have been discussed~\cite{KK_08,CM-SY_08}.  
 On the other hand,  
 a number of groups~\cite{FKSY_94,FKYetal_95,VF_97_arxiv,PS_02} have pursued first-principle derivation 
 of the density functional via effective action.   
  These efforts either introduce an auxiliary field~\cite{FKSY_94,PS_02}  
  or expand in powers of $e^2$~\cite{FKYetal_95,VF_97_arxiv}.
  The strengths of the auxiliary field approach are the simplicity of
  the effective action expression and the fact that each term already includes infinitely many
    Feynman diagrams~\cite{Jackiw_74}.  However, this approach seems~\cite{FKSY_94} to lack a direct connection to 
     the Kohn-Sham (KS) scheme. Such a connection can be made in the expansion in powers of $e^2$~\cite{Sham_85,VF_97_arxiv},
   but that expansion is good only when $e^2$ is small~\cite{Negele_Orland_book}.  The validity of that assumption    
   depends on the strength and variation of $\upsilon(\x)$.

In this \lett, without assuming $e^2$ small, we report our 
development~\cite{yky_prb}
of an auxiliary field method that makes a direct connection to the KS scheme.  
To lighten the mathematical expressions in our finite-temperature formalism, 
 we suppress the spin degree of freedom (as it is easy to include) and 
 denote by a dot (circle) the three (four) dimensional integral contraction (with $\tau$ denoting the Euclidean time, $x\equiv(\tau,\x)$)
\vspace*{-2pt}\phantom{12}
\bea
a \3di b &\equiv & \int d\x \; a(\x)\, b(\x) \nonumber  \vspace*{-4pt}\\
 a \4di b &\equiv& \int\! d\tau d\x \, a(\tau, \x)\, b(\tau, \x) 
 \equiv \int dx \, a(x)\, b(x) \; .  \nonumber 
  \vspace*{-3pt}\phantom{12}
\eea

 To probe the 
 electron density, one introduces  to $\hat H$ a classical source term $J(\x)$
 coupled to $\hpsid(\x) \hpsi(\x)$, $\hat H \to \hat H + J \3di (\hpsid \hpsi) \equiv \hat H_J$. 
 Let $\beta$ be the temperature  inverse, 
 $\beta  J \3di (\hpsid \hpsi)$ is written as $J \4di (\hpsid \hpsi) = \int dx J(x)\hpsid(x)\hpsi(x)$.  
 The partition function now is a functional of $J$, 
 that is 
\be  \label{genfunc.0}
\hspace*{-8pt} Z [J] \Rightarrow e^{-\beta W[J]} \!\!= 
 \Tr \left[ e^{-\beta \left[\hat H + J \3di (\hpsid \hpsi) \right]} \right]
 \! \equiv \! \Tr \left[ e^{-\beta \hat H_{\!J} } \right].
\ee

To disentangle the quartic fermionic interaction, we use the standard procedure
 of introducing an auxiliary field $\phi$  
and express $Z[J]$ as a path integral over both the Grassmann fields and the auxiliary field  \vspace*{-2pt}
\begin{equation}
e^{-\beta W[J]} = 
\int D\phi D\psid D\psi \; e^{ -S\left[ \phi, \psid,\psi \right] } \;,   \label{genfunc}
\vspace*{-4pt}\phantom{12}
\end{equation}
where  \vspace*{-2pt}\phantom{12}
\bea
&& \hspace*{-25pt} S \left[ \phi, \psid \!,\psi \right] =  -\frac{1}{2}\Tr \ln (u) + \frac{1}{2}\phi \4di u \4di \phi 
+ \psid \4di G^{-1} \4di \psi  \label{action.1} \\  
&& \hspace*{-25pt} G^{-1}(x,x')=\left( {\partial \tau } + \hat h(\x)  
 + i (u\4di \phi)_x + J(x) \right)\delta (x-x') \label{GJ} \\
&& \hspace*{-25pt} \hat h(\x)= -\frac{\nabla^2}{2m} + \upsilon_{\rm ion}(\x) - \mu  
\label{s_ptcle_h}  \\ 
&& \hspace*{-25pt} u(x,x') = \delta(\tau-\tau') e^2/|\x-\x'| \equiv \delta(\tau-\tau') u(\x,\x') \label{u.def}\;,\vspace*{-2pt}\phantom{12}
\eea
with $\psi ^{(\dagger) }$ denoting the Grassmann fields satisfying $\psi ^{(\dagger)}(\beta,\x) = -\psi ^{(\dagger)}(0,\x)$.
It is easy to verify that 
\be \label{W2n}
\frac{\delta (\beta W\left[ J \right])}{\delta J(x)} 
 = \la \hpsid(x)\hpsi(x) \ra_J = \la \hat n(x) \ra_J \equiv n_J(x)\; . 
\ee
 Eq.~(\ref{W2n}) expresses $n$ in terms of $J$. The effective action is defined as the Legendre transformation
 of $\beta W[J]$
 \be \label{Gamma.def} 
\Gamma[n_J] \equiv \beta W[J] - J \4di n_J \; ,
 \ee
 where the subscript $J$ indicates that the domain of $\Gamma[n]$ is the set of  density profiles
  reachable by varying $J$. Eq.~(\ref{Gamma.def}) also 
leads to \vspace*{-6pt}
\be \label{Gamma.dn}
\frac{\delta \Gamma[n]}{\delta n} = -J \; .
 \vspace*{-2pt}\phantom{12}
\ee

We now show that $E_\upsilon[n] = \lim_{\beta \to \infty} \frac{1}{\beta} \Gamma[n]$. 
Eq.~(\ref{genfunc.0}) assures that 
 at the zero temperature limit $W[J]$ is simply the ground state energy corresponding
 to $\hat H_J$.  Evidently, when $J=0$, 
 $\lim_{\beta \to \infty} \frac{1}{\beta}\Gamma[n]\vert_{n=n_g} = W[J]|_{J=0} = E_g$
  where $E_g$ stands for 
 the ground state energy corresponding to $\hat H$ and $n_g$ represents the 
 electron density at the physical ($J=0$) ground state. When $J \ne 0$, the 
 corresponding electronic density $n_J$ is different from $n_g$ 
 and $\lim_{\beta \to \infty} \frac{1}{\beta}\Gamma[n]\vert_{n=n_J}$ represents
  the expectation value of $\hat H$, 
  calculated using the ground state wave function corresponding to a different Hamiltonian $\hat H_J$.
 Thus by the definition of the ground state, $\lim_{\beta \to \infty} \frac{1}{\beta}\Gamma[n]|_{n=n_J} > 
 \lim_{\beta \to \infty} \frac{1}{\beta}\Gamma [n]|_{n=n_g}$. 
  This means that 
 $\lim_{\beta \to \infty} \frac{1}{\beta}\Gamma[n]$ reaches its minimum at $n_g$.
 Thus  $\lim_{\beta \to \infty} \frac{1}{\beta}\Gamma[n]$ has all the properties 
 attributed to the energy functional $E_\upsilon$ in (\ref{enfunc}) and (\ref{KSfunctional}).  
 Since the HK theorem states that this functional is unique, it must in fact 
 be equal to $\lim_{\beta \to \infty} \frac{1}{\beta} \Gamma[n]$.

If we make a change of variable $\phi \to \phi + i u^{-1} \4di J$ 
in (\ref{genfunc}-\ref{GJ}) 
 and integrate over the Grassmann fields, we obtain \vspace*{-2pt}
\bea
e^{-\beta W[J]} &\equiv&  e^{\frac{1}2 J \4di u^{-1} \4di J} e^{-\beta W_\phi [J]} \nonumber \\
&=& e^{\frac{1}2 J \4di u^{-1} \4di J} \int D\phi  \; e^{ -I\left[ \phi \right] - iJ\4di \phi } \; ,  \label{genfunc.1}
\vspace*{-2pt}\phantom{12}
\eea
where \vspace*{-2pt}
\be \label{I.phi}
I[\phi] = -\frac{1}2 \Tr \ln (u) + \frac{1}2 \phi \4di u \4di \phi  - \Tr \ln (G_{\phi}^{-1})\, , 
 \vspace*{-2pt}\phantom{1}
\ee
and  \vspace*{-2pt}
\be \label{G.1}
{G}_\phi^{-1}(x,x') 
 =  \left( \partial_\tau +\hat h(\x) + i (u \4di \phi)_x  \right) \delta(x-x') \, . 
  \vspace*{-2pt}\phantom{1}
\ee
Eq.~(\ref{genfunc.1}) implies that  
\be \label{W_phi} 
\beta W[J] = \beta W_{\phi}[J] - \frac{1}{2} J \4di u^{-1} \4di J \, ,
\ee
 and thus the left-hand side of~(\ref{W2n}) can be expressed differently, leading to  
\be \label{n2varphi}
n_J = i\varphi - u^{-1} \4di J \; , 
\ee
where $i\varphi \equiv \delta (\beta W_{\phi}[J])/{\delta J}$.   
 To evaluate $\beta W_\phi$, we follow Jackiw~\cite{Jackiw_74} and let 
 $\phi \to \phi + \varphi$   
 in~(\ref{genfunc.1}-\ref{G.1}). 
 In particular, (\ref{G.1}) is rewritten as  
 \be  \label{G.phi.def}
G_{\phi+\varphi}^{-1}(x,x') = G_\varphi^{-1}(x,x') + i \delta(x-x')
 \left(u\4di \phi \right)_x  \, ,
 \ee
and one obtains~\cite{Jackiw_74}
\bea
&&\hspace*{-28pt}\beta W_\phi [J] = \frac{1}{2}\Tr \ln (\tilde {\cal D}^{-1} \4di u ) + \frac{1}{2} \varphi \4di u \4di \varphi
- \Tr \ln \left( G_{\varphi}^{-1} \right) \nonumber \\
&& \hspace*{-27pt}+ i J \4di \varphi  - \sum_{n=1}^\infty \frac{1}{n!} \la \left[ \sum_{k=3}^\infty
 I^{(k)}[\varphi] \4di \,b_1 \ldots \4di\, b_k  \right]^n \ra_{\rm 1PI,~conn.} , \label{W.fukuda} 
\eea
where the subscript ``${\rm 1PI,~conn.}$" means to include only connected, one-particle-irreducible diagrams,
 $b \equiv u\4di \phi$, \vspace*{-2pt}
\bea
\tilde {\mathcal D}^{-1} &=& u^{-1} - D \; , \nonumber \\
D(x,y) &=& G_\varphi (x,y) G_\varphi(y,x) \; ,  \nonumber 
 \vspace*{-2pt}\phantom{12}
\eea 
and \vspace*{-2pt}
\bea
&& \hspace*{-25pt} I^{(k)}[\varphi] \4di b_1 \ldots \4di b_k  \equiv  
\frac{(-1)^{k-1}}{k} \int \! dx_1 \ldots dx_k \nonumber \\
&& \hspace*{-10pt} G_{\varphi}(x_k,x_1) \ldots G_{\varphi}(x_{k-1},x_k)
 (ib(x_1))\ldots(ib(x_k))\label{I_kernel} \; .  
 \vspace*{-2pt}\phantom{12}
\eea 
Fukuda {\it et al.}~\cite{FKSY_94} obtained an expression similar to (\ref{W.fukuda}) 
and used it to derive an effective action as a functional of $\varphi$. They also noted that
 this auxiliary field approach does not make a direct connection to the KS scheme.

Coming to the point of departure from typical auxiliary field approaches,  
we show below how an exact correspondence to the KS scheme can be made 
 for the auxiliary field method by decomposing the source $J$ in a particular way. 
Let us define a free fermion propagator $\PH_0$ by
\be \label{Green_0}
\PH_0^{-1}(x,x') = \left[ \partial_\tau +\hat h ({\x}) +  J_0(x) \right] \delta (x-x') \; ,  
\ee
where $ J_0$ is chosen (if $\frac{\delta E_{xc}[n]}{\delta n} \vert_{n_J}$ exists, $J_0$ exists
 and can be written~\cite{yky_prb} 
 as $u\3di n_J + \frac{\delta E_{xc}[n]}{\delta n} \vert_{n_J} + J$ ) 
such that  
\be \label{KS_n_free} 
-\PH_0(x,x) = n_J(x) \; . 
\ee
Eq.~(\ref{KS_n_free}) demands that this non-interacting (KS) 
 system have electron density, $-\PH_0(x,x)$, identical to $n_J(x)$, the electronic density of
   the physical system (where Coulomb interactions exist).  In (\ref{W.fukuda}), each occurrence of 
 $i u\4di \varphi$ through $G_\varphi$ is to be replaced by $J + u\4di n_J$ (from (\ref{n2varphi})).

To bring out the KS scheme, we perform
the following source decomposition
\be \label{source.decomp}
J[n] = (J_0[n] - u\4di n_J) + J'[n] \equiv \tilde J_0[n] + J'[n] \; . 
\ee
Then from~(\ref{G.1}) and (\ref{n2varphi}) we have   
\be 
G_\varphi^{-1}(x,x') =  \PH_0^{-1}(x,x') + J'(x) \delta(x-x') \; .
 \label{Green_phi_n}  
\ee

Although the source decomposition (\ref{source.decomp}) is introduced here
 for the first time in the auxiliary field approach,  
  a similar method was used in~\cite{FKYetal_95,VF_97_arxiv} 
 to perform perturbative calculations using 
 $e^2$ as the expansion parameter.

Substituting (\ref{n2varphi}) and (\ref{W.fukuda}) into~(\ref{W_phi}), one
 obtains an expression for $\beta W[J]$, which, upon introducing a parameter $\lambda$
 (to be set $=1$ in the end) to denote the loop order, has the form 
 $\beta W [J] = \beta \tilde W_0 [J] + \sum_{i=1}^\infty \lambda^i (\beta W_i[J+u\4di n_J])$, where in
  particular~\cite{yky_prb} \vspace*{-4pt}
  \be 
\beta \tilde W_0 [J] =  \beta W_0[J+u\4di n_J] - \frac{1}{2} n_J \4di u \4di n_J  \; ,
\vspace*{-2pt}\phantom{12}
  \ee
with $\beta W_0[J+u\4di n_J] = -\Tr \ln (G_\varphi^{-1})$.   

 To arrive at an expansion headed by $-\Tr \ln (\PH_0^{-1})$ instead of 
  $-\Tr \ln (G_\varphi^{-1})$, and containing the expression $W_{l}[J_0]$ instead
   of $W_{l}[J+u\4di n_J]$, we expand $W_l[J + u\4di n_J] = W_l[J_0 + J']$ in powers of $J'$ (subscript $l$ omitted
   in the equation below)\vspace*{-2pt}
\be \label{W.Jp.exp}
W = W[J_0] + \frac{\delta W[J_0]}{\delta J_0} \4di J'
+ \frac{1}{2} \frac{\delta^2 W[J_0]}{\delta J_0\, \delta J_0}
 \4di J' \4di J' + \ldots  
\ee
 The expression $W_l[J_0]$ means that
$J$ is replaced by $\tilde J_0$ but $u\4di n_J$ is  
kept unchanged.\cite{yky_prb} 
With (\ref{W.Jp.exp}), 
we may express $\beta W[J]$ as a double series 
\be \label{W.db} 
 \beta W[J] = \beta \tilde W_{00} + \beta \sum_{i,k} W_{ik}\left(1-\delta_{i,0}\delta_{k,0} \right){J'}^{k} \lambda^i\; ,
\ee
where each $W_{ik}$ involves the $k$'th derivative of $W_i$. In particular, $\tilde W_{00}$ is given
by (with $n_J \to n$ hereafter)  
\be \label{W.00}
\beta \tilde W_{00}  = \beta W_{00} - \frac{1}{2} n \4di u \4di n  = 
 -\Tr \ln (\PH_0^{-1}) - \frac{1}{2} n \4di u \4di n \; ,
\ee
and in view of (\ref{KS_n_free}) $W_{01}$ is given by
\be \label{W0.dJ0}
\frac{\delta (\beta W_0[J_0])}{\delta J_0}   =  n = \frac{\delta (\beta \tilde W_{00}[\tilde J_0])}{\delta \tilde J_0} \; .
\ee
The second half of (\ref{W0.dJ0}) suggests that we define  
\be\label{tGamma.0.def}
\tilde \Gamma_0[n] = \beta \tilde W_{00}[\tilde J_0] - \tilde J_0 \4di n \; ,
\ee
 the Legendre transformation of the zeroth order contribution from $\beta W[J]$ (in terms of $J'$ and $\lambda$),   
leading to 
\be \label{tJ0.def}
\frac{\delta \tilde \Gamma_0[n]}{\delta n} = -\tilde J_0 \; .
\ee

Comparing (\ref{tJ0.def}) with (\ref{Gamma.dn}), 
 we find \vspace*{-2pt}
 \be \label{Gamma.int.der}
\frac{\delta (\Gamma[n]- \tilde \Gamma_0[n])}{\delta n} = -J'  \;.
\vspace*{-2pt} \phantom{12}
 \ee
The idea now is to develop a series for $\Gamma[n]$ led by $\tilde \Gamma_0[n]$. 
 Subtracting (\ref{tGamma.0.def}) from (\ref{Gamma.def}), 
 we have \vspace*{-1pt}
\be \label{Gamma.int.def}
\Gamma[n] - \tilde \Gamma_0[n] = \beta W[J] - \beta \tilde W_{00}[\tilde J_0] - J' \4di n \; ,
\vspace*{-1pt} \phantom{12}
\ee
in which the last two terms on the right hand side exactly cancel the terms in $\tilde W_{00}$ and
$W_{01}$ contributing to $\beta W[J]$. So the series for $\Gamma - \tilde \Gamma_0$ is just 
(\ref{W.db}) with those two terms removed. Next we convert the double sum in (\ref{W.db}) 
 into a single sum by expanding $J'$ as a series in $\lambda$. We write  \vspace*{-2pt}
 \be \label{Jp.def}
 J'[n] = \sum_{l=1}^\infty J_l [n] \lambda^l \; ,
 \vspace*{-2pt} \phantom{12}
 \ee
where the precise expressions for $J_1, J_2,\ldots$ are as yet undetermined since (\ref{Jp.def})
is not a loop expansion. We substitute (\ref{Jp.def}) formally into (\ref{Gamma.int.def}) and (\ref{W.db}) 
to obtain a series \vspace*{-4pt} \phantom{12} 
\be \label{Gamma.loop}
\Gamma[n]- \tilde \Gamma_0[n] = \sum_{l=1}^\infty \Gamma_l[n] \; \lambda^l \; , 
\vspace*{-2pt} \phantom{12}
\ee
in which each $\Gamma_l$ is defined explicitly in terms of the $J_k$, $\beta W_{k \le l}[J_0]$,
and their derivatives.  
Because $W_{01}$ is missing from (\ref{Gamma.int.def}), any occurrence of $J_k$ is accompanied by at
least one other factor $J_{k'}$ or else by an occurrence of some $W_{i>0}$, and hence by
a power of $\lambda$ higher than the $k$'th. In other words, the expression for $\Gamma_{l \ge 1}$ involves only $J_k$ with $ k < l$. We finally remove the indeterminacy in (\ref{Jp.def}) by imposing 
 (\ref{Gamma.int.der}) to hold order by order in $\lambda$, leading to  
 \be \label{J_l} 
 \frac{\delta \Gamma_l[n]}{\delta n} = - J_l \; .
 \ee

Since $\Gamma_{l\ge 1}$ involves only $J_{k < l}$,  all
the $J_l$ and $\Gamma_l$ can be found explicitly by applying (\ref{Gamma.loop}) and (\ref{J_l}) alternately. The first few expressions are 
$\Gamma_1 = \beta W_1[J_0] = -\frac{1}{2} \Tr \ln (\tilde {\mathcal D}_{\!\!J \to \tilde J_0}^{-1} \4di u )$, $J_1 = -\frac{\delta (\beta W_1[J_0])}{\delta J_0} 
\4di \frac{\delta J_0}{\delta n}$, $\Gamma_2 = \beta W_2[J_0] 
 + \frac{\delta (\beta W_1[J_0])}{\delta J_0} \4di J_1
  + \frac{1}{2} J_1 \4di \frac{\delta^2 (\beta W_0[J_0])}{\delta J_0 \delta  J_0}
   \4di J_1$. \vspace*{5pt}
  
For an arbitrary $J_0$, one will obtain a corresponding density $\tilde n$. The computation of $\frac{1}{\beta}\Gamma[n]$
using (\ref{tGamma.0.def}), (\ref{Gamma.loop}) and (\ref{J_l}) 
evaluates the energy functional at density $\tilde n$, which may or may not be the 
 ground state density. To obtain the ground state density and the corresponding $J_0$, one needs to 
  solve at zero temperature limit the extremal equation $ 0 = \frac{\delta \Gamma [n]}{\delta n}$, which we turn to shortly.      

 To carry out the calculation of $\Gamma[n]$, we need to compute $J_l$ (see (\ref{J_l})) via
 the functional derivative 
\be \label{dc.def}   
\frac{\delta}{\delta n} = \left( \frac{\delta n}{\delta J_0}\right)^{-1} \4di \frac{\delta}{\delta J_0}
 \, \equiv\,  D_0^{-1} \4di \frac{\delta}{\delta J_0} \; . 
\ee  
Diagrams corresponding to $\beta W_l[J_0]$ and their derivatives  
 contain the $u$, $\PH_0$, and $\tilde {\mathcal D}_0 \equiv 
 \tilde {\mathcal D}_{\!\!J\to \tilde J_0}$ propagators. 
It is easy to show that one may express $\delta n(x)/\delta J_0(y)$ as
\be 
- \frac{\delta \PH_0(x,x)}{\delta J_0(y)} = 
  \PH_0(x,y) \, \PH_0(y,x) = D_{J\to \tilde J_0}(x,y) \label{dndJ.tdep}
\ee 
and thus 
$D_0^{-1} = D_{J\to \tilde J_0}^{-1}$, which we call the inverse density correlator. 
The differentiation rules of $\PH_0$, $\tilde {\mathcal D}_0$, and $D_0^{-1}$ with respect to $J_0$
 can be expressed diagrammatically: 
\[
\begin{picture}(215,40)(10,0)
\Line[dash,dashsize=2.5](10,30)(45,30) \Vertex(10,30){1.0} \Vertex(45,30){1.0}
\Line[double](135,30)(170,30) \Vertex(135,30){1.0} \Vertex(170,30){1.0}
\Line[arrow,arrowlength=3.0,arrowwidth=0.8,arrowinset=0.2](10,10)(45,10) 
\Vertex(10,10){1.0} \Vertex(45,10){1.0}
\Photon[](135,10)(170,10){2}{9} \Vertex(135.5,10){1.8} \Vertex(169.5,10){1.8} 
\Text(65,30)[lc]{$u(x,x')$}
\Text(-2,28)[lb]{$x'$} \Text(57,28)[rb]{$x$}
\Text(190,30)[lc]{$D_0^{-1}(x,x')$}
\Text(123,28)[lb]{$x'$} \Text(182,28)[rb]{$x$}
\Text(65,10)[lc]{${\mathcal G}_0(x,x')$}
\Text(-2,8)[lb]{$x'$} \Text(57,8)[rb]{$x$}
\Text(123,8)[lb]{$x'$} \Text(182,8)[rb]{$x$}
\Text(192,10)[lc]{$\tilde {\mathcal D}_0 (x,x')$}
\end{picture} 
\] 
\vspace*{-40pt}
\bea 
\frac{\delta \PH_0(x,x') }{\delta J_0(y)}
 = \frac{\delta }{\delta  J_0(y)} \; 
\begin{picture}(10,40)(0,-3)
\Line[arrow,arrowlength=3.0,arrowwidth=0.8,arrowinset=0.2](5,-20)(5,20)
\Vertex(5,-20){1.0}
\Vertex(5,20){1.0}
\Text(5,-24)[tc]{$x'$}
\Text(5,24)[bc]{$x$}
\end{picture}
&=& - \; \begin{picture}(10,40)(0,-3)
\Line[arrow,arrowlength=3.0,arrowwidth=0.8,arrowinset=0.2](5,-20)(5,0)
\Line[arrow,arrowlength=3.0,arrowwidth=0.8,arrowinset=0.2](5,0)(5,20)
\Vertex(5,0){1.8}
\Vertex(5,-20){1.0}
\Vertex(5,20){1.0}
\Text(5,-24)[tc]{$x'$}
\Text(5,24)[bc]{$x$}
\Text(11,-4)[bc]{$y$}
\end{picture} \nonumber \\
\frac{\delta \tilde{\mathcal D}_0(x,x') }{\delta  J_0(y)} = 
\frac{\delta }{\delta J_0(y)} \; 
\begin{picture}(10,55)(0,-3)
\Photon[](5,-20)(5,20){2}{9} 
\Vertex(5,-20){1.8}
\Vertex(5,20){1.8}
\Text(5,-24)[tc]{$x'$}
\Text(5,24)[bc]{$x$}
\end{picture}
&=&  - \; \begin{picture}(30,40)(0,-3)
\Photon[](15,-20)(15,-10){2}{3}
\Photon[](15,10)(15,20){2}{3}
\Arc[arrow,arrowpos=0.50,arrowlength=3.0,arrowwidth=0.8,arrowinset=0.2,clock](15,0)(10,-90,-180)
\Arc[arrow,arrowpos=0.50,arrowlength=3.0,arrowwidth=0.8,arrowinset=0.2,clock](15,0)(10,-180,-270)
\Arc[arrow,arrowpos=0.50,arrowlength=3.0,arrowwidth=0.8,arrowinset=0.2,clock](15,0)(10,90,-90)
\Vertex(5,0){1.8}
\Vertex(15,-20){1.8}
\Vertex(15,20){1.8}
\Vertex(15,-10){1.8}
\Vertex(15,10){1.8}
\Text(15,-24)[tc]{$x'$}
\Text(15,24)[bc]{$x$}
\Text(11,-4)[bc]{$y$}
\end{picture}
\; - \;  \begin{picture}(30,40)(0,-3)
\Photon[](15,-20)(15,-10){2}{3}
\Photon[](15,10)(15,20){2}{3}
\Arc[arrow,arrowpos=0.50,arrowlength=3.0,arrowwidth=0.8,arrowinset=0.2,clock,flip](15,0)(10,-90,-180)
\Arc[arrow,arrowpos=0.50,arrowlength=3.0,arrowwidth=0.8,arrowinset=0.2,clock,flip](15,0)(10,-180,-270)
\Arc[arrow,arrowpos=0.50,arrowlength=3.0,arrowwidth=0.8,arrowinset=0.2,clock,flip](15,0)(10,90,-90)
\Vertex(5,0){1.8}
\Vertex(15,-20){1.8}
\Vertex(15,20){1.8}
\Vertex(15,-10){1.8}
\Vertex(15,10){1.8}
\Text(15,-24)[tc]{$x'$}
\Text(15,24)[bc]{$x$}
\Text(11,-4)[bc]{$y$}
\end{picture} \nonumber \\
\frac{\delta D_0^{-1}(x,x') }{\delta J_0(y)} 
= \frac{\delta }{\delta J_0(y)} \; 
\begin{picture}(10,55)(0,-3)
\Line[double](5,-20)(5,20)
\Vertex(5,-19.5){1.0}
\Vertex(5,19.5){1.0}
\Text(5,-24)[tc]{$x'$}
\Text(5,24)[bc]{$x$}
\end{picture}
&=&  + \; \begin{picture}(30,40)(0,-3)
\Line[double](15,-20)(15,-10) \Vertex(15,-19.5){1.0}
\Line[double](15,10)(15,20) \Vertex(15,19.5){1.0}
\Arc[arrow,arrowpos=0.50,arrowlength=3.0,arrowwidth=0.8,arrowinset=0.2,clock](15,0)(10,-90,-180)
\Arc[arrow,arrowpos=0.50,arrowlength=3.0,arrowwidth=0.8,arrowinset=0.2,clock](15,0)(10,-180,-270)
\Arc[arrow,arrowpos=0.50,arrowlength=3.0,arrowwidth=0.8,arrowinset=0.2,clock](15,0)(10,90,-90)
\Vertex(5,0){1.8}
\Vertex(15,10){1.5}\Vertex(15,-10){1.0}
\Text(15,-24)[tc]{$x'$}
\Text(15,24)[bc]{$x$}
\Text(11,-4)[bc]{$y$}
\end{picture}
\; + \;  \begin{picture}(30,40)(0,-3)
\Line[double](15,-20)(15,-10) \Vertex(15,-19.5){1.0}
\Line[double](15,10)(15,20) \Vertex(15,19.5){1.0}
\Arc[arrow,arrowpos=0.50,arrowlength=3.0,arrowwidth=0.8,arrowinset=0.2,clock,flip](15,0)(10,-90,-180)
\Arc[arrow,arrowpos=0.50,arrowlength=3.0,arrowwidth=0.8,arrowinset=0.2,clock,flip](15,0)(10,-180,-270)
\Arc[arrow,arrowpos=0.50,arrowlength=3.0,arrowwidth=0.8,arrowinset=0.2,clock,flip](15,0)(10,90,-90)
\Vertex(5,0){1.8}
\Vertex(15,10){1.0}\Vertex(15,-10){1.0}
\Text(15,-24)[tc]{$x'$}
\Text(15,24)[bc]{$x$}
\Text(11,-4)[bc]{$y$}
\end{picture} \; .\nonumber  \\
&& \nonumber
\eea
\vspace*{8pt}
 
 The differentiation rules of $\PH_0$, $\tilde {\mathcal D}_0$, and $D_0^{-1}$ with respect to $n$ are simply obtained by compounding the
  results above with (\ref{dc.def}). We show only one example:
 \vspace*{-8pt}
\[
\frac{\delta \PH_0(x,x') }{\delta n(z)}
 = \frac{\delta }{\delta n(z)} \; 
\begin{picture}(10,40)(0,-3)
\Line[arrow,arrowlength=3.0,arrowwidth=0.8,arrowinset=0.2](5,-20)(5,20)
\Vertex(5,-20){1.0}
\Vertex(5,20){1.0}
\Text(5,-24)[tc]{$x'$}
\Text(5,24)[bc]{$x$}
\end{picture}
\; =\;  - \; \begin{picture}(30,40)(-20,-3)
\Line[arrow,arrowlength=3.0,arrowwidth=0.8,arrowinset=0.2](5,-20)(5,0)
\Line[arrow,arrowlength=3.0,arrowwidth=0.8,arrowinset=0.2](5,0)(5,20)
\Line[double](-15,0)(5,0)
\Vertex(-15,0){1.8}
\Vertex(5,0){1.0}
\Vertex(5,-20){1.0}
\Vertex(5,20){1.0}
\Text(5,-24)[tc]{$x'$}
\Text(5,24)[bc]{$x$}
\Text(-15,-4)[tc]{$z$}
\end{picture} \; .
\]
\vspace*{20pt}

Equipped with these differentiation rules, one may use standard diagrammatic expansion to
compute the $W_l[J_0]$s, their functional derivatives with respect to
$J_0$, as well as $J_l$s to facilitate the calculations of $\Gamma_l$s. Because 
$D_0(x,y) = \PH_0(x,y)\PH_0(y,x)$, both $\tilde {\mathcal D}_0
  = \left( u^{-1} -D_0 \right)^{-1}$ and $D_0^{-1}$ can be expressed in terms of single-particle orbitals and energies 
  through $\PH_0(x,y)$ --the propagator of the KS system-- which can be expressed as 
\bea
\PH_0(x,y) &=& \sum_\alpha \phi_\alpha(\x) \phi_\alpha^*(\y) e^{-(\ve_\alpha-\mu)(\tau_x - \tau_y)} \times \nonumber \\
&& \times  \left\{ \begin{array}{l r}(-n_\alpha) & {\rm if~} \tau_x \le \tau_y \\
(1-n_\alpha) & {\rm if~} \tau_x > \tau_y \end{array} \right. \; , \label{Green.orbital} 
\eea
where $n_\alpha = 1/(e^{\beta (\ve_\alpha-\mu)} +1)$, $\sum_\alpha n_\alpha =  N_e$, and 
the single particle orbital $\phi_\alpha(\x)$ satisfies 
\[
\left[ \hat h(\x) + J_0(\x) \right] \phi_\alpha(\x) = (\ve_\alpha - \mu) \phi_\alpha(\x) \; . 
\]

Since $\frac{\delta \tilde \Gamma_0[n]}{\delta n} = -\tilde J_0$,  the extremal condition $ 0 = \frac{\delta \Gamma[n]}{\delta n} $
 that determines $n_g$ and $J_0[n_g]$ (as $\beta \to \infty$) becomes 
\be \label{eq:self.consistent} 
\frac{\delta \left(\sum_{i=1}^\infty \Gamma_i[n]\right)}{\delta n}  = 
D_0^{-1} \4di \frac{\delta \left( \sum_{i=1}^\infty \Gamma_i [J_0[n]] \right)}{\delta J_0} =  \tilde J_0 \; .
\ee 
Eq.~(\ref{eq:self.consistent}) has to be solved self-consistently by keeping $\Gamma_i$ terms up to some order in $\lambda$. 
 Although a truncation is necessary, we note that each diagram in our expression already corresponds to infinitely many Feynman 
  diagrams when using $e^2$ as the expansion parameter. This is easily seen by performing the
   small $e^2$ expansion of  $\tilde {\mathcal D}_0$ 
 \[
\tilde {\mathcal D}_0 = u + u\4di D_0 \4di u +  u\4di D_0 \4di u \4di D_0 \4di u + \ldots \;, 
 \]     
a sum of infinitely many (dressed) propagators. Interestingly, in the strong coupling limit where one must treat $u^{-1}$ as
 a small parameter, we may express $\tilde {\mathcal D}_0$ as  
\[
\tilde {\mathcal D}_0  = -D_0^{-1} - D_0^{-1} \4di u^{-1} \4di D_0^{-1} 
- D_0^{-1} \4di u^{-1} \4di D_0^{-1}\4di u^{-1} \4di D_0^{-1} - \ldots  
\] 
while the traditional $e^2$ expansion fails completely.

Finally we sketch how (\ref{KSfunctional}) arises from $\Gamma[n]$. Eq.~(\ref{tGamma.0.def}) may be rewritten as
$\frac{1}{\beta} \tilde \Gamma_0[n] = \frac{1}{\beta}\left[ - \Tr \ln (\PH_0^{-1} ) - J_0 \4di n \right] +  \frac{1}{2\beta} n\4di u \4di n$. 
Because  $- \Tr \ln (\PH_0^{-1} ) = \sum_\alpha  \ln (1-n_\alpha)$,
 at zero temperature limit, the first two terms of $\frac{1}{\beta} \tilde\Gamma_0$ above give rise to the $T_0[n] - \mu N_e + \int \upsilon (\x) n(x) d\x$
  while the last part is 
  exactly the Hartree term~\cite{yky_prb}. The exchange-correlation functional $E_{xc}[n]$ equals $\lim_{\beta \to \infty} 
  \frac{1}{\beta} \sum_{i=1}^\infty \Gamma_i[n]$.   
We also comment that the excitations of the system can be studied~\cite{yky_prb} under this formalism 
 and the energy functional shown in this letter has the correct single-electron limit~\cite{yky_prb}.

 Providing a scheme beyond perturbative expansion in $e^2$, we have proposed an effective action
  construction that will contribute to the development of the parameter-free universal density functional.

This research was supported by the Intramural Research Program of the National Library of Medicine of
 the National Institutes of Health. 
 \vspace*{-12pt} 

\vspace*{-12pt}
\end{document}